# Strain-tuning for superconductivity in La$_3$Ni$_2$O$_7$ thin films

Motoki Osada[1,2,*], Chieko Terakura[3], Akiko Kikkawa[3], Masamichi Nakajima[3], Hsiao-Yi Chen[4], Yusuke Nomura[4], Yoshinori Tokura[2,3,5], Atsushi Tsukazaki[1,2,4,*]

[1] Quantum-Phase Electronics Center (QPEC), The University of Tokyo, Hongo, Tokyo, Japan
[2] Department of Applied Physics, The University of Tokyo, Hongo, Tokyo, Japan
[3] RIKEN Center for Emergent Matter Science (CEMS), Wako, Saitama, Japan
[4] Institute for Materials Research (IMR), Tohoku University, Sendai, Miyagi, Japan
[5] Tokyo College, The University of Tokyo, Hongo, Tokyo, Japan

## Abstract

The recent discovery of high-transition temperature ($T_c$) superconductivity in pressurized La$_3$Ni$_2$O$_7$ bulk crystals has attracted keen attention for its characteristic energy diagram of $e_g$ orbitals containing nearly half-filled $d_{3z^2-r^2}$ and quarter-filled $d_{x^2-y^2}$ orbitals. This finding provides valuable insights into the orbital contributions and interlayer interactions in double NiO$_6$ octahedrons that further provides a chance to control the electronic structure via varying ligand field. Here, we demonstrate that strain-tuning of the $T_c$ over a range of 50 K with La$_3$Ni$_2$O$_7$ films on different oxide substrates under 20 GPa. As the $c/a$ ratio increases, the onset $T_c$ systematically increases from 10 K in the tensile-strained film on SrTiO$_3$ to the highest value about 60 K in the compressively strained film on LaAlO$_3$. These systematic variations suggest that strain-engineering is a promising approach for expanding the superconductivity in bilayer nickelates with tuning the energy diagram for achieving high-$T_c$ superconductivity.

________________________________________________
osada@ap.t.u-tokyo.ac.jp, tsukazaki@ap.t.u-tokyo.ac.jp

## Introduction

The recent discovery of high-$T_c$ superconductivity near 80 K in pressurized La$_3$Ni$_2$O$_7$ crystals[1] has sparked renewed interest in the broader family of nickelate compounds[2–12]. This nickelate belongs to the Ruddlesden-Popper (RP) series ($RE_{n+1}$Ni$_n$O$_{3n+1}$, $RE$ = rare-earth element), consisting of bilayer NiO$_6$ octahedra separated by a $RE$ rock-salt layer, with one example being La$_3$Ni$_2$O$_7$ shown in Fig. 1a. Surprisingly, signatures of superconductivity at ambient pressure have been also reported in strained thin films[13,14]. In a simplified picture of the electronic band structure, Ni $3d$ orbital adopts a $d^{7.5}$ electron configuration. Density functional theory (DFT) calculations suggest that the $d_{3z^2-r^2}$ orbital facilitates significant out-of-plane hopping $t_\perp$ via apical oxygens of the NiO$_6$ octahedra[15,16,17]. This interlayer interaction causes the $d_{3z^2-r^2}$ bands to split into anti-bonding and bonding states, with the centroid of the $d_{x^2-y^2}$ band placed in between (Fig. 1a). In this bilayer Hubbard model, the $d_{3z^2-r^2}$ and $d_{x^2-y^2}$ orbitals are nearly half-filled and quarter-filled, respectively, and the strong coupling between the layers through the $d_{3z^2-r^2}$ orbital is believed to play a critical role for the emergence of high-$T_c$ superconductivity. This bilayer and multi-orbital nature provide a distinct perspective from the high-$T_c$ cuprates, where the single-layer and single-orbital model with nearly half-filled $d_{x^2-y^2}$ orbital becomes a good description[18].

In the studies based on bulk crystals, it has been discussed that structural phase transition triggers the superconducting transition at roughly 10–14 GPa (Ref.[4,5]); an orthorhombic *Amam* phase at ambient pressure transforms to a tetragonal *Fmmm* or *I4/mmm* phase[1,5,19], which enhances interlayer coupling essential for the electronic structure preferred for high-$T_c$ superconductivity. Another structural factor is the possible involvement of monolayer and trilayer structural units in contributing to superconductivity[20,21,22]. Such structural stability or uniformity is largely tied to oxygen deficiencies and the coexistence of competing RP phases ($n$ = 1 and 3), which can hinder the optimization of superconducting properties. Specifically, oxygen vacancies, especially at inner apical oxygen sites, disrupt $d_{3z^2-r^2}$ orbital hybridization that is crucial for facilitating superconductivity, often inducing insulating behavior instead[23,24]. To address these issues, concerted efforts in advancement in material synthesis are essential for achieving robust high-$T_c$ superconductivity in this emerging class of nickelates. In this study, we applied a combination of thin-film fabrication technique based on coherent epitaxy and measurement under hydrostatic pressure for strain-tuning of the lattice and electronic structure. Strain tuning of oxide thin films offers large possibility to control the symmetry of octahedron in perovskite structures, resulting in the emergence of fascinating phenomena, for example variation of superconducting properties of cuprates[25,26], emergence of ferroelectricity in SrTiO$_3$ (Ref.[27]) and metal-insulator transition in manganites[28]. In the structural point of view, the ligand field of octahedra is directly linked to the lattice distortion, which likely provides an opportunity to control the energy diagram of the bilayer nickelates. The intuitive scenario under a compressive situation, with a large $c/a$ ratio elongating the NiO$_6$ octahedra along $c$-axis, is as follows: the energy separation ($\Delta E$) between $d_{x^2-y^2}$ and $d_{3z^2-r^2}$ orbitals in the energy diagram shown in Fig. 1a is expected to increase, due to ascending the energy level of the $d_{x^2-y^2}$ orbital and lowering the $d_{3z^2-r^2}$ orbital. Such modifications in the ligand field could effectively tune the electronic band structure and orbital filling, offering a promising route to maximize the superconducting properties in bilayer nickelates.

**Epitaxial thin film**

We fabricated La$_3$Ni$_2$O$_7$ thin films on three substrates using pulsed-laser deposition (see Methods for details). To tune the strain, we selected SrTiO$_3$ (001) with ($a_{STO}$ = 3.905 Å), NdGaO$_3$ (001)$_{pc}$ ($a_{NGO}$ = 3.855 Å) and LaAlO$_3$ (001) ($a_{LAO}$ = 3.787 Å) as substrates. While NdGaO$_3$ adopts an orthorhombic *Pbnm* structure, here we employ a pseudocubic perovskite unit cell for the following discussion. Under ambient pressure, bulk La$_3$Ni$_2$O$_7$ adopts an orthorhombic *Amam* structure with lattice parameters $a$ = 5.412 Å, $b$ = 5.456 Å, and $c$ = 20.45 Å. This structure can be approximated as a pseudo-tetragonal unit cell with $a_{pt}$ = $b_{pt}$ ~ 3.833 Å and $c$ = 20.45 Å. Accordingly, the applicable strain in films is summarized in Fig. 1b. The film on SrTiO$_3$ imposes a tensile strain of +1.9%, NdGaO$_3$ close to the bulk lattice but still introduces a slight tensile strain of +0.6 %, and that on LaAlO$_3$ is a compressive strain of –1.2%. Figure 1c shows the high-angle annular dark-field (HAADF) scanning transmission electron microscopy (STEM) cross-sectional image of La$_3$Ni$_2$O$_7$ film on NdGaO$_3$ substrate, which exhibits high-quality film with abrupt interface. While additional NiO$_6$ layers were sparsely detected as 3-NiO$_6$ layers, corresponding to approximately 6% off-stoichiometry, the bilayer structural units were predominantly observed, separated by LaO rock-salt layers, corresponding to the 2222-structure (see also Supplementary Information Fig. S1 for an overview and interface).

The crystalline structure was characterized using x-ray diffraction techniques. Judging from no additional peak except for La$_3$Ni$_2$O$_7$ and substrate materials in 2theta-omega scans for three films, a single phase of $La_3Ni_2O_7$ is fabricated on the substrates (see Supplementary Information Fig. S2). Crucially, we conducted phi-scans around the $La_3Ni_2O_7$ (2 0 6) peaks to investigate the crystal rotation symmetry. The phi-scans revealed clear four-fold symmetry of the (2 0 6) peaks with 45-degree rotation from the substrate (1 0 1) peaks (see Supplementary Information Fig. S3). This rotation symmetry and appearance of a single peak in reciprocal space mapping (RSM) discussed below strongly suggests that the $La_3Ni_2O_7$ thin films are stabilized in the tetragonal symmetry owing to epitaxial strain. RSMs were measured around the substrate (1 0 3) peak, as shown in Fig. 1d for $LaAlO_3$ (left) and $NdGaO_3$ (right) (A cross-point of two red broken lines indicates bulk values of $La_3Ni_2O_7$). In both RSMs, a peak corresponding to $La_3Ni_2O_7$ (1 1 $\underline{17}$) was aligned at substrate along $a$-axis direction, indicating that the films are almost strained to the in-plane lattice of substrates. The in-plane lattice constant was evaluated to be 3.815±0.014 Å for $La_3Ni_2O_7$ on $LaAlO_3$ and 3.855 Å for that on $NdGaO_3$, corresponding to −0.5% and +0.6% strains, respectively. Compared with the perfect matching of the film lattice to that of $NdGaO_3$, the strain of the film on $LaAlO_3$ is partially relaxed. The experimentally obtained lattice constants are summarized in Fig. 1e, compared with reported bulk values (open diamonds: $La_3Ni_2O_7$ bulk crystal under ambient pressure and after structural phase transition at 14 GPa (Ref.[1])). The out-of-plane lattice constant correspondingly exhibits elongation on $LaAlO_3$ ($c$ = 20.58 Å) and contraction on $NdGaO_3$ ($c$ = 20.41 Å). Please note that the $a$-axis value for the film on $SrTiO_3$ is plotted to top axis due to difficulty of experimental evaluation with weak diffraction intensity in RSM measurement. Considering the Poisson's ratio ($\nu$ = 0.5), both films are well-strained in-plane with small variation in the out-of-plane lattices, exhibiting minimal oxygen deficiency.

Based on this lattice condition at ambient pressure, we estimate the lattice distortion under hydrostatic pressure. The compression ratio of tetragonal phase of $La_3Ni_2O_7$ bulk single crystal around hydrostatic pressure $P$ = 20−40 GPa is approximately $-3.22 \times 10^{-3}$ Å/GPa in pseudo-tetragonal estimation for averaged $a,b$-axis and $-16.6 \times 10^{-3}$ Å/GPa for $c$-axis[1]. Considering the tetragonal lattice unit, hydrostatic pressure compresses the lattice of the $La_3Ni_2O_7$ crystal almost isotropically. As shown in Supplementary Information Fig. S4, the compression ratio of substrate materials with perovskite structure that was calculated using DFT-based methods (see Methods and Supplementary Information for details) is close values in $-(5 \sim 6.5) \times 10^{-3}$ Å/GPa. By applying $-3.22 \times 10^{-3}$ Å/GPa, we roughly estimate the compressed $a$-axis length of $La_3Ni_2O_7$ films as shown in Fig. 1f. When assuming the critical lattice constant for a structural phase transition or the appearance of superconductivity, a higher critical pressure is required to reach it in tensile-strained films with a larger $a$-axis length. It is also an intriguing viewpoint to compare the critical lattice constant with that for the strained film exhibiting superconductivity at ambient pressure[13]. In the future, direct measurement of the lattice constant of the films under applied pressure using X-ray diffraction with a diamond anvil cell may become feasible.

## Superconductivity in $La_3Ni_2O_7$ films under hydrostatic pressure

We measured the electrical resistivity of three films under applying hydrostatic pressure up to 20 GPa using a cubic anvil cell with a liquid pressure-transmitting medium. In Fig. 2a, the superconducting transition was clearly observed at $P = 20$ GPa for the films on three substrates. The onset $T_c$ and the temperature at resistivity reaching zero ($T_c^{zero}$) for the film on $LaAlO_3$ are approximately 60 and 48 K, respectively. The value of $T_c^{zero}$ is comparable to the reported bulk values of 40 K (Ref.[4]) and 41 K (Ref.[6]). It should be noted that the resistivity of the film on $NdGaO_3$ and $LaAlO_3$ at normal state shows a linear-in-temperature, indicating a strange metal phase. In addition, the onset $T_c$ of the films on $NdGaO_3$ and $SrTiO_3$ decreases to 40 and 10 K, respectively, which is in accord with the increase of the residual resistance. Figures 2b–f exhibit temperature dependent resistivity $\rho_{xx}$ ($T$) of $La_3Ni_2O_7$ films on $SrTiO_3$ (Fig. 2b), $NdGaO_3$ (Fig. 2c and d), and $LaAlO_3$ (Fig. 2e and f). At $P = 0$ GPa, while the film on $SrTiO_3$ shows metallic behaviour (Fig. 2b), the other two are insulating (Fig. 2c and e). Metallic behaviour has been also observed in bulk samples[1,5]. At low pressure regions roughly below 12 GPa, the insulating behaviour is pronounced in all films, which is possibly related to the formation of spin or charge density wave phase as suggested by previous studies on bulk samples[23,29,30,31]. We also note that local variations of oxygen content in thin film might be close situation as recently reported in bulk samples exhibiting such insulating phases under pressure. We focus on the clear kink of resistivity that correspond to the characteristic temperature for the formation of the density wave phase as $T_{DW}$, at $P = 1$ GPa, appears at 170 K for the film on $SrTiO_3$, 175 K for that on $NdGaO_3$, and 150 K for that on $LaAlO_3$. The kink at $T_{DW}$ decreases and becomes broader with increasing pressure, indicating suppression of the density wave phase formation. By further increasing the pressure, all films exhibited metallic $\rho_{xx}-T$ curves, resulting in superconducting phase (Fig. 2b, d and f). Current−voltage characteristics were carried out within the current upper limit of $10^{-4}$ A. Although finite voltage generation was observed with approximately $10^{-5}$ A in low-$T_c$ film on NdGaO$_3$ (see Supplementary Information Fig. S5), the high-$T_c$ film on LaAlO$_3$ would require higher currents than our measurement setup. Notably, the large $c/a$ with elongated $c$-axis and compressed $a$-axis, that is La$_3$Ni$_2$O$_7$/LaAlO$_3$, stabilizes superconductivity, highlighting the critical role of strain engineering in maximizing superconducting properties.

**Phase diagram**

Figures 3a–c presents the temperature–pressure ($T$–$P$) phase diagrams of La$_3$Ni$_2$O$_7$ films on three substrates, which was based on the temperature dependence of resistivity shown in Fig. 2. Yellow, purple, and blue represent the metallic, insulating, and superconducting regions, respectively. First, the insulating density wave phase is strongly suppressed in the film on SrTiO$_3$. However, weak insulating behaviour appears at below approximately 50 K (purple), resulting in narrow region of superconductivity in the $T-P$ diagram. Second, the superconducting phase appears in close proximity to insulating state, indicating that the suppression of density wave phase is a key to the phase transition. This behavior is reminiscent of features associated with high-$T_c$ superconductivity in cuprates[18,32], suggesting a correlation between the insulating and superconducting phases. Third, the superconducting region with the resistance reaching zero appears in the films on NdGaO$_3$ and LaAlO$_3$. The rich phases in the $T-P$ diagrams of three films indicate that the stain effect effectively induces variations in phase stability.

**Strain effect**

The strain effect for the electrical properties in three films are discussed with the critical pressure ($P_c$) for appearance of superconductivity (Fig. 4a), $T_{DW}$ (Fig. 4b), and $T_c$ (Fig. 4c) as a function of $c/a$. It has been discussed that hydrostatic pressure plays two roles in the appearance of superconductivity in bulk studies[1,4,5]: one is inducing the structural phase transition, and the other is increasing the hybridization of electronic bands. The strain effect in thin films is significantly different from the isotropic compression under applying hydrostatic pressure, because the increase in the $c/a$ in the film corresponds to the compression of the $a$-axis length and elongation of the $c$-axis length, or vice versa. Although the $c/a$ ratio varies in range of approximately 6% in three films, the $P_c$ seems almost constant about 12−16 GPa, being reasonable in comparison with that of bulk values. This result indicates that the structural phase transition is not a major trigger to induce the superconductivity in La$_3$Ni$_2$O$_7$. The degree of hybridization may be a dominant factor in the films. Since no structural phase transition is observed in thin films under strain, while maintaining the tetragonal symmetry, it is plausible that hydrostatic pressure enhances hybridization via compression. Although we did not directly measure the structural evolution or hybridization under hydrostatic pressure in this study, the correlation between strain- and pressure-induced changes and the superconducting phase diagram suggests that hybridization plays an important role. Spectroscopic studies with a diamond anvil cell will provide further physical insight to directly clarify the orbital evolution under pressure.

Two characteristic temperatures $T_{DW}$ at $P = 1$ GPa and $T_c$ at 20 GPa are summarized in Fig. 4b and c. Comparing the values of the films on NdGaO$_3$ and LaAlO$_3$, a decrease in $T_{DW}$ and an increase in $T_c$ are observed with an increasing the $c/a$ ratio. This suggests that the change in the ligand field and the increase of $\Delta E$ is closely related to the stabilization of superconductivity. In contrast, the $T_{DW}$ for SrTiO$_3$ is not on the trend probably due to metallic background conduction. The electron conduction in La$_3$Ni$_2$O$_{7-\delta}$ bulks has been reported to undergo a spin-order phase at around 150 K and a charge-order phase at approximately 115 K (Ref. [23,24,29,30,31,33–36]). In particular, the role of oxygen deficiencies was discussed on the formation of these density wave phases in previous bulk studies[23,29,30,31]. Although it is generally difficult to quantitatively analyze oxygen deficiencies in thin films, the effectiveness of ozone annealing to induce metallic behaviour and superconductivity have been reported in thin-film samples[13,14]. Considering the suppression of insulating behaviour in the La$_3$Ni$_2$O$_7$ film on LaAlO$_3$ (see Supplementary Information Fig. S7) after ozone annealing, the observed $T_{DW}$ in the as-grown films with tetragonal structure is likely linked to the oxygen deficiencies, which may be as comparable to the reported conditions in both thin films[13] and bulk samples[19]. Based on the good quality of the film, it should be pronounced that the film with large $c/a$ on LaAlO$_3$ show superior superconducting properties with the highest $T_c^{zero}$ around 50 K. The systematic increase of $T_c$ demonstrates that the strain tuning enables us to effectively control the $\Delta E$ in energy diagram for the emergence of superconductivity.

## Conclusion

We have demonstrated the systematic strain tuning of the superconducting transition temperature in La$_3$Ni$_2$O$_7$ thin films based on the coherent epitaxy technique. By applying this approach, we successfully achieved the highest $T_c^{zero} = 48$ K in the compressively strained film on LaAlO$_3$.

Notably, we found that the critical pressure for superconductivity remains unchanged within this lattice tuning regime. Our results about the lattice engineering in terms of *c*/*a* will contribute to construct a unified picture of the superconductivity in bilayer nickelates under isotropic compression by hydrostatic pressure and anisotropic variation by epitaxial strain. Further investigations into strain tuning of RP nickelates are warranted[37–40], particularly for suppressing competing phases, tuning the electronic band structure, and optimizing the superconducting gap, all of which are critical for achieving high-$T_c$ superconductivity.

## Methods

**Sample preparation:** La$_3$Ni$_2$O$_7$ films were fabricated on single-crystalline SrTiO$_3$ (001), NdGaO$_3$ (001)$_{pc}$, and LaAlO$_3$ (001) substrates using pulsed-laser deposition. La$_3$Ni$_2$O$_7$ polycrystalline targets were ablated by a KrF excimer laser (wavelength 248 nm). Substrates were pre-annealed at 750°C in an oxygen partial pressure of 1 × 10$^{-6}$ Torr to obtain an atomically flat surface. During growth, the substrate temperature was fixed at 650°C and the oxygen partial pressure was 200 mTorr. The laser fluence was ~1.0 J/cm$^2$ and a repetition rate was 4 Hz. The film thickness was in a range of 10 to 20 nm by adjusting the laser pulse counts. The films were characterized using x-ray diffraction (XRD) techniques with Cu K$\alpha$ source ($\lambda$ = 1.5406 Å). The interface quality of La$_3$Ni$_2$O$_7$/NdGaO$_3$ (001)$_{pc}$ was examined using scanning transmission electron microscopy (STEM). Crystal structures were visualized using the VESTA software[41].

**Device for high-pressure measurement:** The thin-film samples were cut into dimensions of 500–1000 μm in lateral size, and the substrates were mechanically polished to reduce their thickness to 0.3 mm (see Supplementary Information Fig. S6 for details). Gold wires were bonded to the samples in a van der Pauw geometry using silver paste to ensure reliable electrical contacts. The $\rho_{xx}$-$T$ curves at $P$ = 0 GPa were observed in physical properties measurement system (PPMS) (Quantum Design, Inc.).

**High-pressure electric resistivity measurement:** The measurements of the temperature dependent resistivity $\rho_{xx}$ under high-pressure was performed under various hydrostatic pressure of 1–20 GPa using a cubic anvil cell with a liquid pressure-transmitting medium (see Supplementary Information Fig. S7 for resistivity variation under applied pressure at $T$ = 300 K). The temperature dependence of resistance was measured using Keithley 2182A nanovoltmeters and Keithley 6221 source meters in delta mode. Measurements were conducted over a temperature range from 292 K to 4.2 K.

**Theoretical calculations based on density functional theory:** The first-principles DFT calculations for structural optimization are performed using Quantum Espresso[42] with a pressure convergence threshold of ±10$^{-3}$ GPa. The calculations adopt the scalar-relativistic PAW (projector augmented wave) pseudopotentials[43] generated by the PSLibrary package[44], for which the exchange-correlation potential is treated within the GGA-PBEsol approximation[45]. The computational parameters are detailed in Table S1 where the lattice constants were converged within an error of ±0.002 angstrom for all target materials.

## Data availability

The data that support the findings of this study are available from the corresponding author upon request.

## Acknowledgements

The authors thank M. Kawasaki with fruitful discussion. STEM observations were made with the cooperation of Y. Kodama and T. Konno of Analytical Research Core for Advanced Materials, Institute for Materials Research, Tohoku University. A part of this work was supported by Tohoku University in MEXT Advanced Research Infrastructure for Materials and Nanotechnology in Japan (Grant No. JPMXP1224TU0193), Basic Science Research Projects by The Sumitomo Foundation, Toyota Riken Scholar Program by Toyota Physical and Chemical Research Institute, and The Kazuchika Okura Memorial Foundation. Y.N. acknowledges support from MEXT as "Program for Promoting Researches on the Supercomputer Fugaku" (Grant No. JPMXP1020230411), JSPS KAKENHI (Grant Nos. JP23H04869, JP23H04519, and JP23K03307), and JST (Grant No. JPMJPF2221).

## Author contributions

M.O. and A.T. conceived the project. M.O. fabricated and characterized nickelate films. M.O., C.T., A.K., M.N., Y.T., and A.T. performed the high-pressure measurement and analysis. H.-Y.C. and Y.N. conducted theoretical calculation. M.O. and A.T. wrote the manuscript with input from all the authors.

## Competing interests

The authors declare no competing interests.

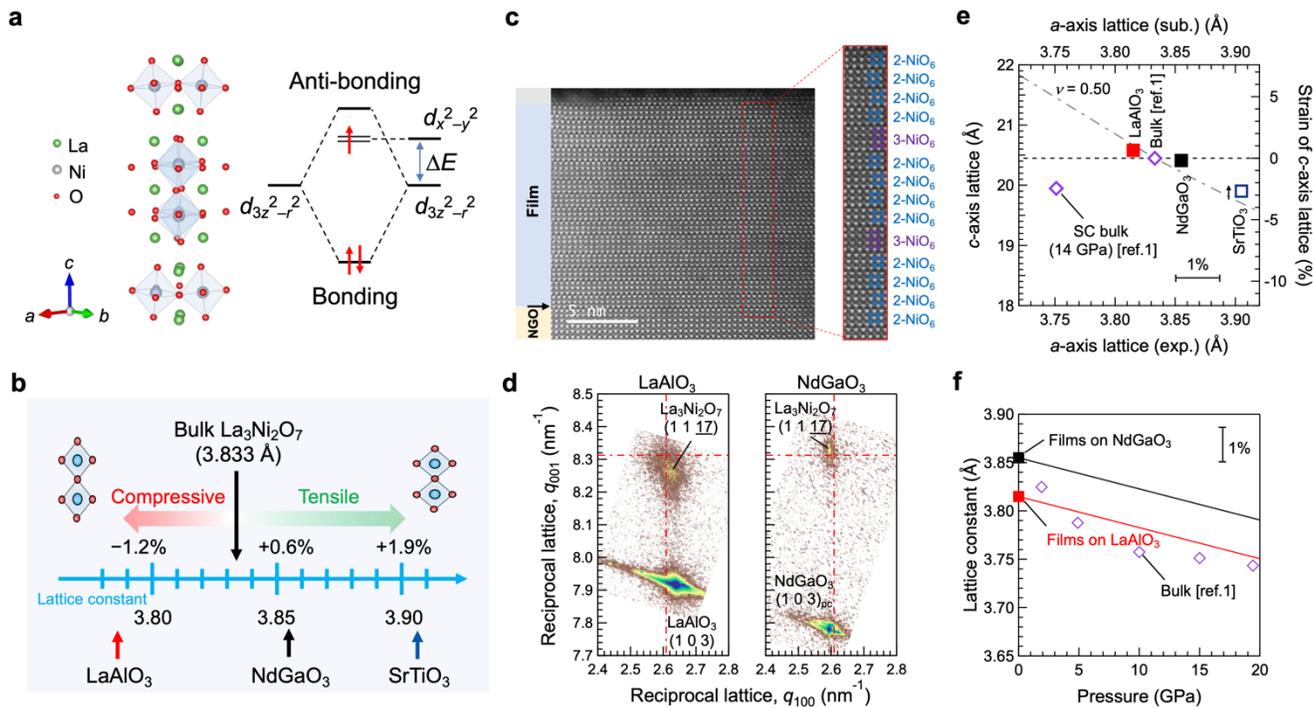

**Fig. 1 | Strain-tuning for ligand field control of La$_3$Ni$_2$O$_7$ thin films. a,** Crystal structure of bilayer nickelates La$_3$Ni$_2$O$_7$ and the energy diagram of Ni $e_g$ orbitals. **b,** Lattice constants of SrTiO$_3$, NdGaO$_3$, and LaAlO$_3$ substrates. **c,** HAADF-STEM cross-sectional image of a La$_3$Ni$_2$O$_7$ thin film grown on NdGaO$_3$ taken along the La$_3$Ni$_2$O$_7$ [110] axis (scale bar 5 nm). Arrow represents the interface. **d,** Reciprocal space maps of La$_3$Ni$_2$O$_7$ (11$\underline{17}$)/ LaAlO$_3$ (103) (left) and La$_3$Ni$_2$O$_7$ (11$\underline{17}$)/NdGaO$_3$ (103) (right). **e,** Experimentally obtained lattice constants of La$_3$Ni$_2$O$_7$ films grown on SrTiO$_3$, NdGaO$_3$, and LaAlO$_3$ substrates. Bulk values are shown in open diamonds[1]. **f.** Calculations of $a$-axis length of La$_3$Ni$_2$O$_7$ films on NdGaO$_3$ and LaAlO$_3$ by assumption of bulk compression ratio. Bulk values are plotted with open diamonds[1].

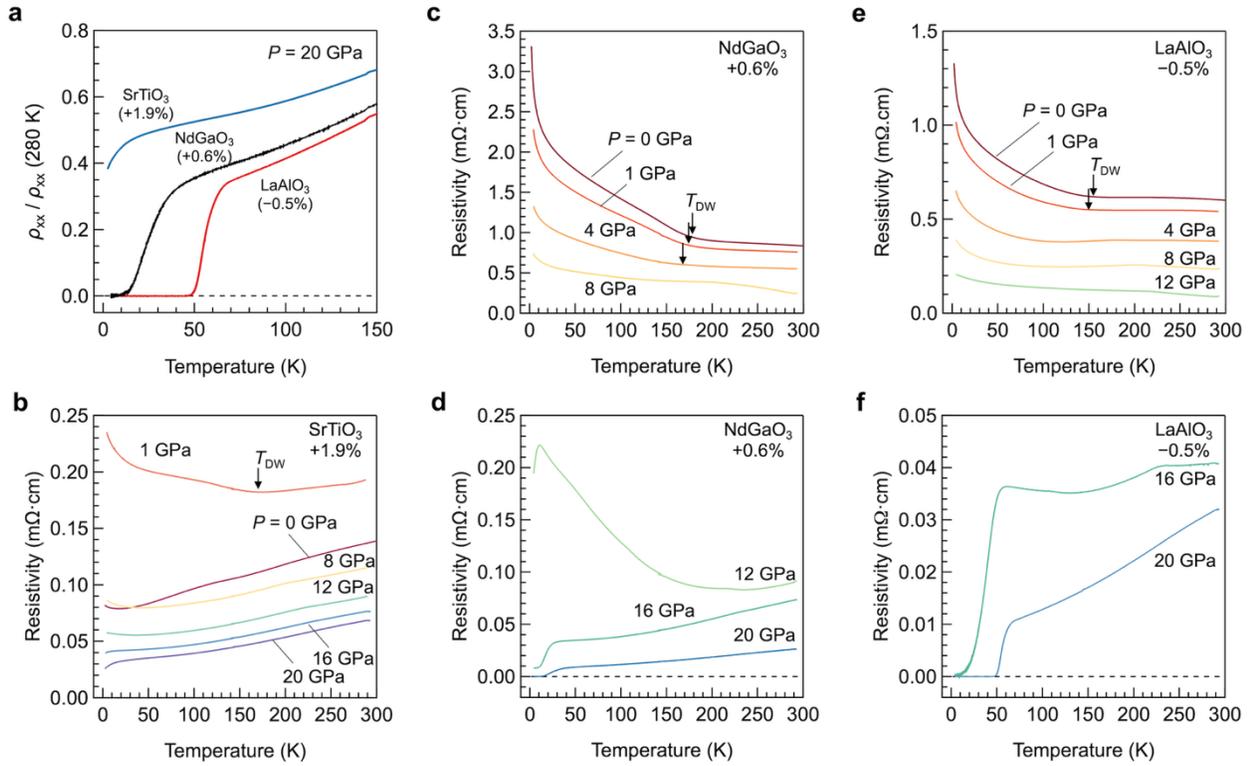

**Fig. 2 | Pressure-induced superconducting transition in temperature dependence of electrical resistivity of La$_3$Ni$_2$O$_7$ films. a,** Substrate dependence of normalized resistivities as a function of temperature under $P$ = 20 GPa. Resistivities are normalized by values at 280 K. **b,** Resistivities as a function of temperature under various pressure for La$_3$Ni$_2$O$_7$ films on SrTiO$_3$. **c,** Resistivities as a function of temperature under $P$ = 0–8 GPa and **d,** those under $P$ = 12–20 GPa for La$_3$Ni$_2$O$_7$ films on NdGaO$_3$. **e,** Resistivities as a function of temperature under $P$ = 0–12 GPa and **f,** those under $P$ = 16–20 GPa for La$_3$Ni$_2$O$_7$ films on LaAlO$_3$.

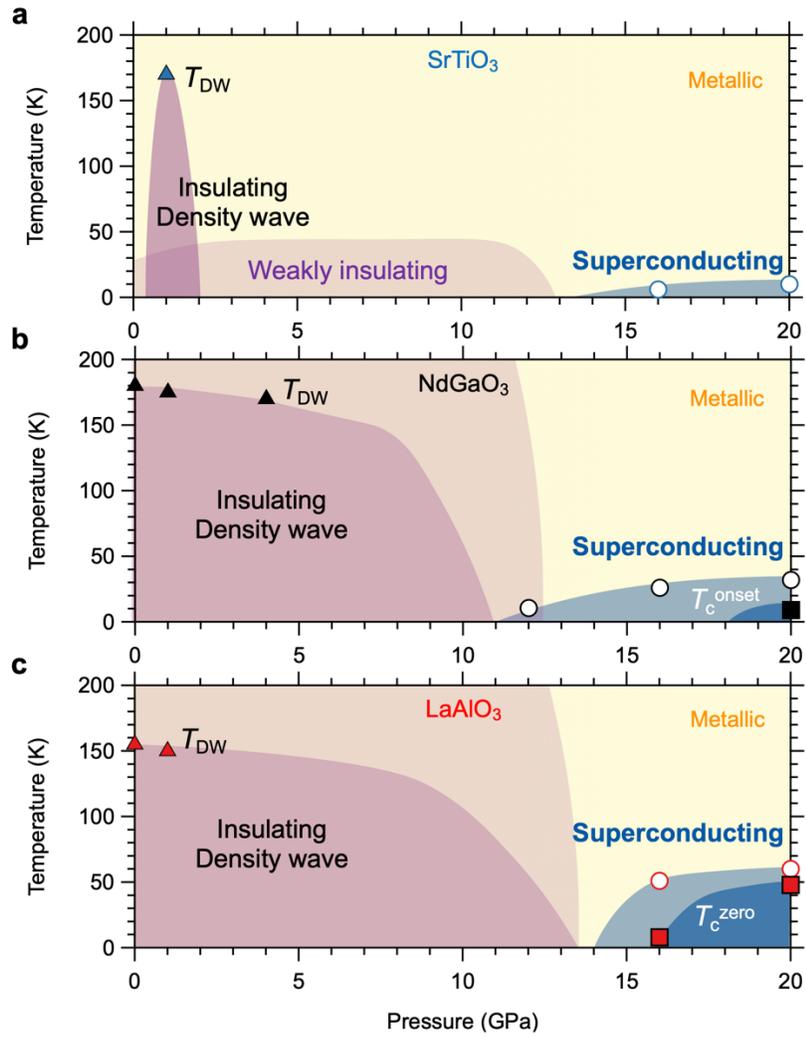

**Fig. 3 | Phase diagrams of $La_3Ni_2O_7$ films.** Temperature–Pressure phase diagrams for $La_3Ni_2O_7$ films on **a,** $SrTiO_3$, **b,** $NdGaO_3$, and **c,** $LaAlO_3$ substates. $T_{DW}$ (closed triangles) is defined by temperatures that resistivity changes discontinuously. $T_c^{zero}$ (close squares) and $T_c^{onset}$ (open circles) are defined by temperatures of onset of resistivity drops and zero-resistance.

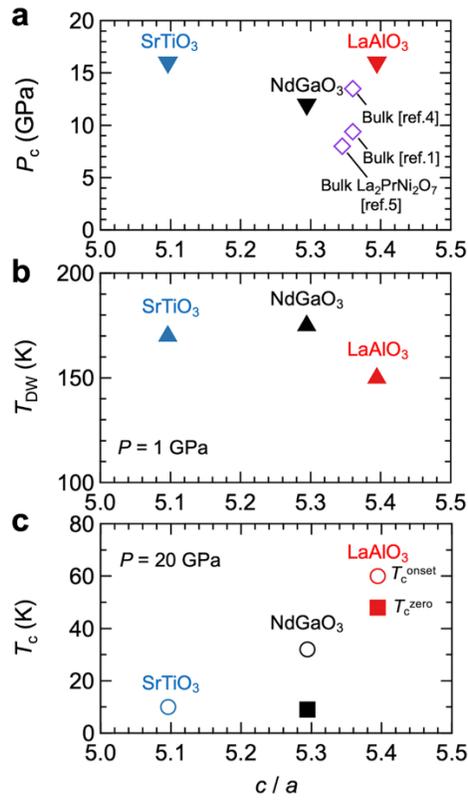

**Fig. 4 | Strain tuning for appearance of superconductivity.** Comparison as a function of $c/a$ ratio. **a,** Critical pressure $P_c$ for appearance of superconductivity. Bulk values are plotted as open diamonds[1,4,5]. **b,** Temperatures for the formation of density wave phase $T_{DW}$. **c,** Critical temperatures of $T_c^{zero}$ (close squares) and $T_c^{onset}$ (open circles).